\documentclass[preprint,12pt]{elsarticle}

\usepackage{amsmath}    
\usepackage{graphicx}   
\usepackage{verbatim}   
\usepackage{color}      
\usepackage{subfigure}  
\usepackage{hyperref}   

\journal{Nuclear Inst. and Methods in Physics Research, A}
\begin{document}
\begin{frontmatter}
\title{Determining the Drift Time of Charge Carriers in P-Type Point-Contact HPGe Detectors}

\author[lbl]{R. D. Martin \corref{cor1}}
\author[lbl]{M. Amman}
\author[lbl]{Y.D. Chan}
\author[lbl]{J.A. Detwiler}
\author[lbl]{J.C. Loach}
\author[lbl,ucb]{Q. Looker}
\author[lbl]{P.N. Luke}
\author[lbl]{A.W.P. Poon}
\author[lbl]{J. Qian\fnref{private}}
\author[lbl,ucb]{K. Vetter}
\author[lbl]{H. Yaver}

\cortext[cor1]{Corresponding author, rdmartin@lbl.gov}
\address[lbl]{Nuclear Science Division, Lawrence Berkeley National Laboratory, 1 Cyclotron Road, Berkeley, CA 94720, USA}
\address[ucb]{Department of Nuclear Engineering, University of California, 4155 Etcheverry Hall, MC 1730, Berkeley, CA 94720, USA}
\fntext[private]{Present Address: Schlumberger-Doll Research, Cambrigde, MA 02139-1578, USA}

\date{\today}
\begin{abstract}

An algorithm to measure the drift time of charge carriers in p-type point contact (PPC) high-purity germanium (HPGe) detectors from the signals processed with a charge-sensitive preamplifier is introduced. It is demonstrated that the drift times can be used to estimate the distance of charge depositions from the point contact and to characterize losses due to charge trapping. A correction for charge trapping effects over a wide range of energies is implemented using the measured drift times and is shown to improve the energy resolution by up to 30\%.

\end{abstract}


\end{frontmatter}

\section{Introduction}
P-type point contact (PPC) high-purity germanium (HPGe) detectors \cite{luke_89, Barbeau:2007qi} have excellent properties for use in research in astro-particle and nuclear physics. They have generated substantial interests in neutrinoless double-beta decay searches by the {\sc Majorana Demonstrator} \cite{MJGeneral:2010i} and GERDA collaborations \cite{Agostini:2010ke} due to their ability to distinguish between multiple and single-site interactions \cite{Budjas:2009zu,Agostini:2010rd}. The weighting potential in these detectors is sharply peaked near the point contact, giving rise to distinguishable current pulses for simultaneous charge depositions separated by as little as 1\,mm. Background from multiple-site charge depositions, such as gamma rays scattering in the detector, are thus distinguishable from neutrinoless double-beta decays which deposit their energy in a small volume.

Another feature of these detectors is that the low capacitance of the small, point-like, collecting contact results in very low levels of electronic noise. This allows the use of PPC detectors with energy thresholds below 0.5\,keV and has made them attractive candidates for direct dark matter detection \cite{Barbeau:2007qi, Aalseth:2010vx}.

The weaker fields in a PPC detector present some challenges. For example, determining the exact start time of the drift of charge carriers can be difficult as the signal generally originates in a region of weak weighting potential where it may be obscured by noise. The drift time is dependent on the location of the ionizing interaction in the crystal; measuring it can thus be a way to monitor the distribution of energy depositions, which could be of use in identifying sources of background in an experiment. Another challenge is the longer drift times in PPC detectors which can result in a degradation of the energy resolution due to charge trapping by impurities in the crystal. It has been shown \cite{kephart_09} that the drift time of events can be used to correct the measured energy, resulting in a substantial improvement in energy resolution.

This paper introduces a digital algorithm to improve the determination of the drift time of events using the rising time of the charge pulses. The performance of this method is then investigated using simulated pulse shapes from energy depositions at different points in a detector. It is shown how the drift times can be used to obtain an estimate of the distance between charge depositions and the point contact and how this is approximated by their z-position (along the longitudinal axis of the crystal). Additionally, it is shown how the charge trapping in a detector can be examined and how the correction introduced in \cite{kephart_09} can be improved by using an optimized measure of drift time.

\section{Pulse drift time measurements and validation}
\label{sec:math}
PPC detectors have a sharply localized weighting potential around the point contact. Charges thus induce most of the signal on the electrode when they are closest to the point contact. An induced pulse will generally have 2 components, a slow component when the charges are far from the electrode and a fast component when the charges are close to the electrode. Traditional rise time metrics (such as the time between the 10\% and 90\% points of the charge pulse, as used in \cite{kephart_09}) are thus biased towards measuring the time that charge carriers spend close to the electrode rather than the actual drift time.

\begin{figure}[!hb]
   \centering
\subfigure{\includegraphics[width=0.45 \textwidth]{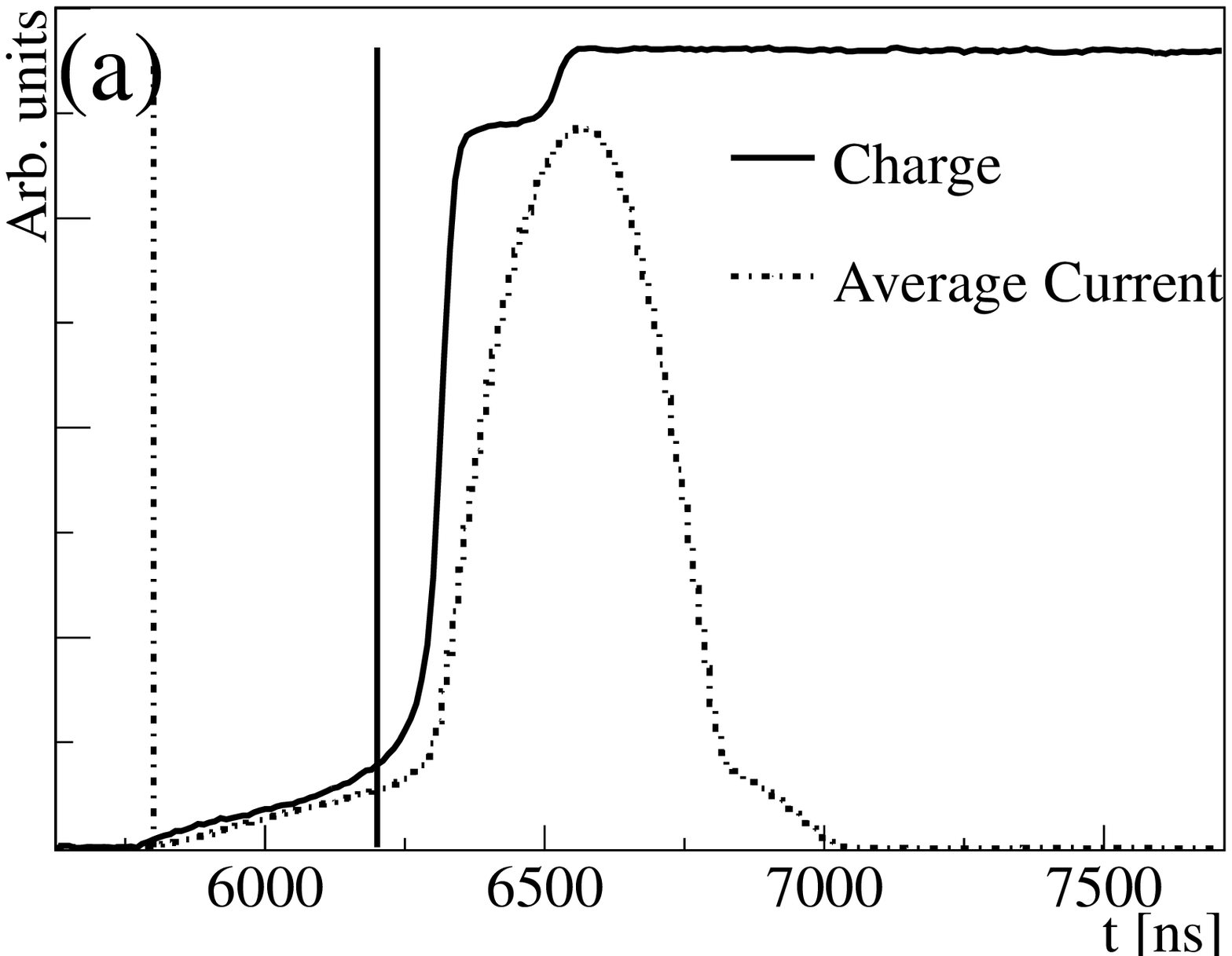}}
\subfigure{\includegraphics[width=0.45 \textwidth]{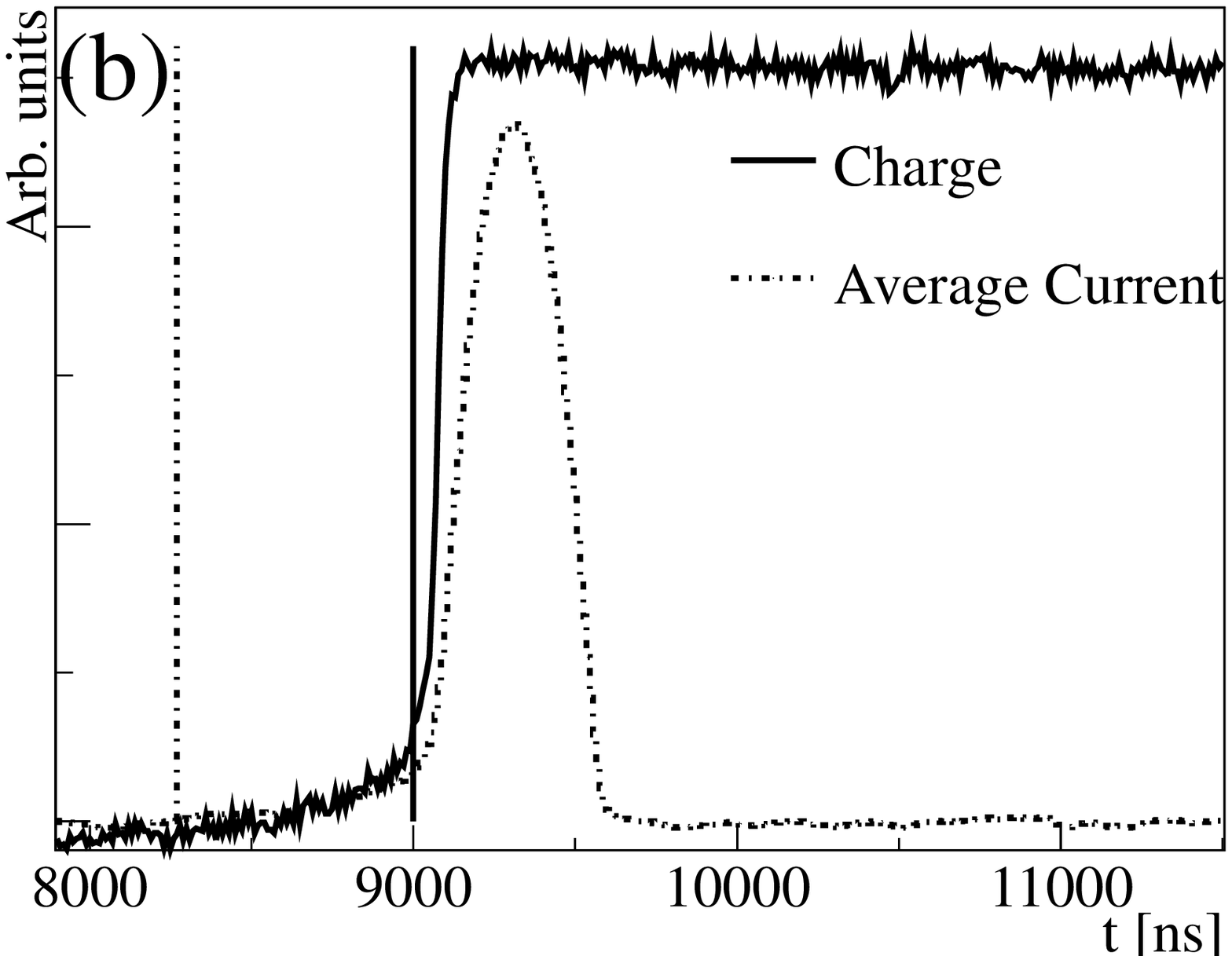}}
\caption{\label{fig:currentrise} Charge pulse (solid line) with average current pulse (dashed line) for a 691\,keV (panel a) and 58\,keV (panel b) event. The vertical lines show,  $t_{10}$, the 10\% rise time of the waveform (solid line) and, $\tilde{t}_{0.1}$, the 0.1\% start time of the current (dashed line). The traditional $t_{10}$ misses much of the actual drift time in the initial slow component. Using the charge pulse to determine a start time is difficult because of noise, as seen in the lower energy event. The high energy event is a characteristic `multi-site' event, likely from a gamma ray interacting twice in the crystal.}
\end{figure}

Figure \ref{fig:currentrise} shows a comparison of the start time, $t_{10}$, defined as the 10\% rise of the charge pulse (solid) and, $\tilde{t}_{0.1}$, defined as the 0.1\% rise of the average current pulse (dashed). Events of two different energies (691\,keV and 58\,keV) are shown. For the higher energy event (panel (a)), the charge waveform could be used to determine an earlier start time; however, this would be much more difficult to do for the lower energy waveform because of noise. One should also note that, in both cases, using $t_{10}$ misses much of the slow component of the pulses, which is most directly correlated to the time that charges drift in the crystal. 

\begin{figure}[!h]
   \centering
\includegraphics[width=0.7 \textwidth]{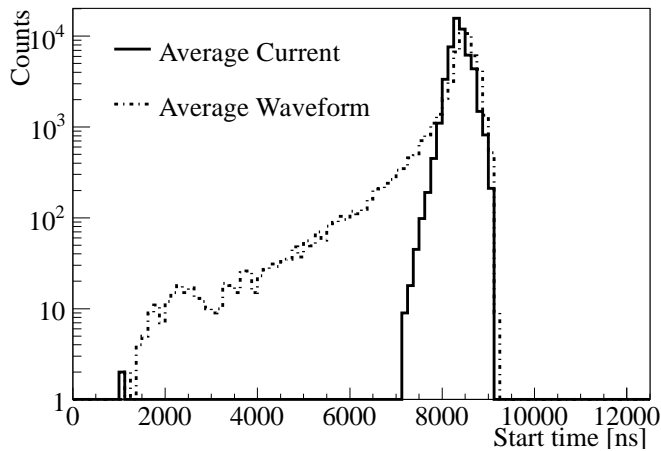}
\caption{\label{fig:TStart_CVsAW_60keV} Comparison of the distributions of starting times determined as the 0.1\% rise of the average current (solid) and average waveform (dashed). The average current and waveform were both determined using a 500\,ns averaging window and the algorithm in equation \ref{eqn:algo}. The data are energy depositions between 58\,keV and 62\,keV collected with the custom point contact detector described in section \ref{sec:data}. There is an apparent bias for the waveform-based calculation to determine an (unphysical) earlier start time.}
\end{figure}

 The difficulty in determining an accurate pulse start time lies in removing biases introduced by noise. Different methods were explored for removing the noise without biasing the start time and it was found that the average derivative of the pulse resulted in the least bias (see Figure \ref{fig:TStart_CVsAW_60keV}, discussed below). Methods to de-noise the waveform, such as a Daubechies wavelet-based filter or a running average were found to introduce distortions in the initial rising edge of the waveform that resulted in a small bias towards earlier start times.

A filtering algorithm that performs a running linear regression to a discretely sampled signal, developed for various PPC detector pulse-processing applications, was found to be useful here in determining the average derivative of a pulse for use in the drift time measurement. Given a set of samples from a digitized charge waveform, $q_i$, collected from the output of a charge-sensitive preamplifier at discrete times $t_i$ and an averaging window length, $W$, the algorithm performs the linear regression, $o_i +s_i i$, to the signal in the time window from $i$ to $i+W$. This effectively gives the slope, $s_i$, and the offset, $o_i$, of a linear fit to the signal between $i$ and $i+W$. The slope and offset for each value of $i$ were calculated as:

\begin{align}
\label{eqn:algo}
\langle y_i\rangle &\equiv \sum_{k=i}^{i+W}q_k \,\,\,\,\,\,\,\,
\langle x_i\rangle \equiv \sum_{k=i}^{i+W}k \nonumber \\
\langle x^2_i\rangle &\equiv \sum_{k=i}^{i+W}k^2 \,\,\,\,\,\,\,\,
\langle xy_i\rangle  \equiv \sum_{k=i}^{i+W}k\,q_k \nonumber \\
\nonumber \\
s_i &= \frac{W\langle xy_i\rangle - \langle x_i\rangle \langle y_i\rangle}{W\langle x^2_i\rangle - \langle x_i\rangle \langle x_i\rangle}\nonumber\\
o_i &= \frac{\langle y_i\rangle -s_i \langle x_i\rangle}{W}
\end{align}

The pulse defined as the set of $s_i$ is the desired average derivative of the charge signal over $W$ samples, and is defined herein as the average current pulse. Additionally, the algorithm also gives the average waveform, defined by the set of $\langle y_i\rangle$, integrated over the same number of samples. $\tilde{t}_{0.1}$, the point at which the averaged current reaches 0.1\% of its maximum value (determined by moving backwards from the maximum) was found to be a good point for defining the start of the drift time. An optimal value for the averaging window was determined to be 500\,ns in the apparatus discussed later in this paper. This was found to be most sensitive to the slow component of the charge pulse without introducing any significant bias. The end time of the drift, $t_{90}$, can be taken as the point where the charge pulse reaches 90\% of its maximum.

Figure \ref{fig:TStart_CVsAW_60keV} shows a comparison of the distributions of start times determined using the average current and the average waveform. In both cases, the start time was determined as the point when the average current (or average waveform) decreased to 0.1\% of its maximum value (stepping backwards in time), with a 500\,ns averaging window.  The data collected correspond to events between 58\,keV and 62\,keV, where noise makes it difficult to determine the starting time of the drift (see Figure \ref{fig:currentrise}). There is a clear bias for the average waveform-based calculation to give earlier rise times that did not correlate with the known position of the events in the crystal.

\section{Monte-Carlo validation of drift time measurement}

A point contact detector with the same geometry and charge impurity profile as the detector used in the measurements (section \ref{sec:data}) was simulated. The detector has a radius of 30.75\,mm and a height of 50\,mm. The simulation was based on code developed by the GRETINA collaboration \cite{Lee2004255} and is described in \cite{siggen,Cooper2011303}. Charge pulses from the detector were simulated for energy depositions on a 2\,mm x 2\,mm grid without including the electronics chain or noise. Gaussian noise, based on the amplitude of the pulse, was later added to the simulated pulses with amplitudes consistent with that seen in the data (1\% noise for a 60\,keV event, 0.1\% noise for a 1\,MeV event, relative to the maximum amplitude of the pulses). The drift time measuring algorithm was then applied to the simulated pulses so that a comparison with the drift time obtained from the simulation directly could be made.

\begin{figure}[!h]
   \centering
\subfigure{\includegraphics[width=0.45 \textwidth]{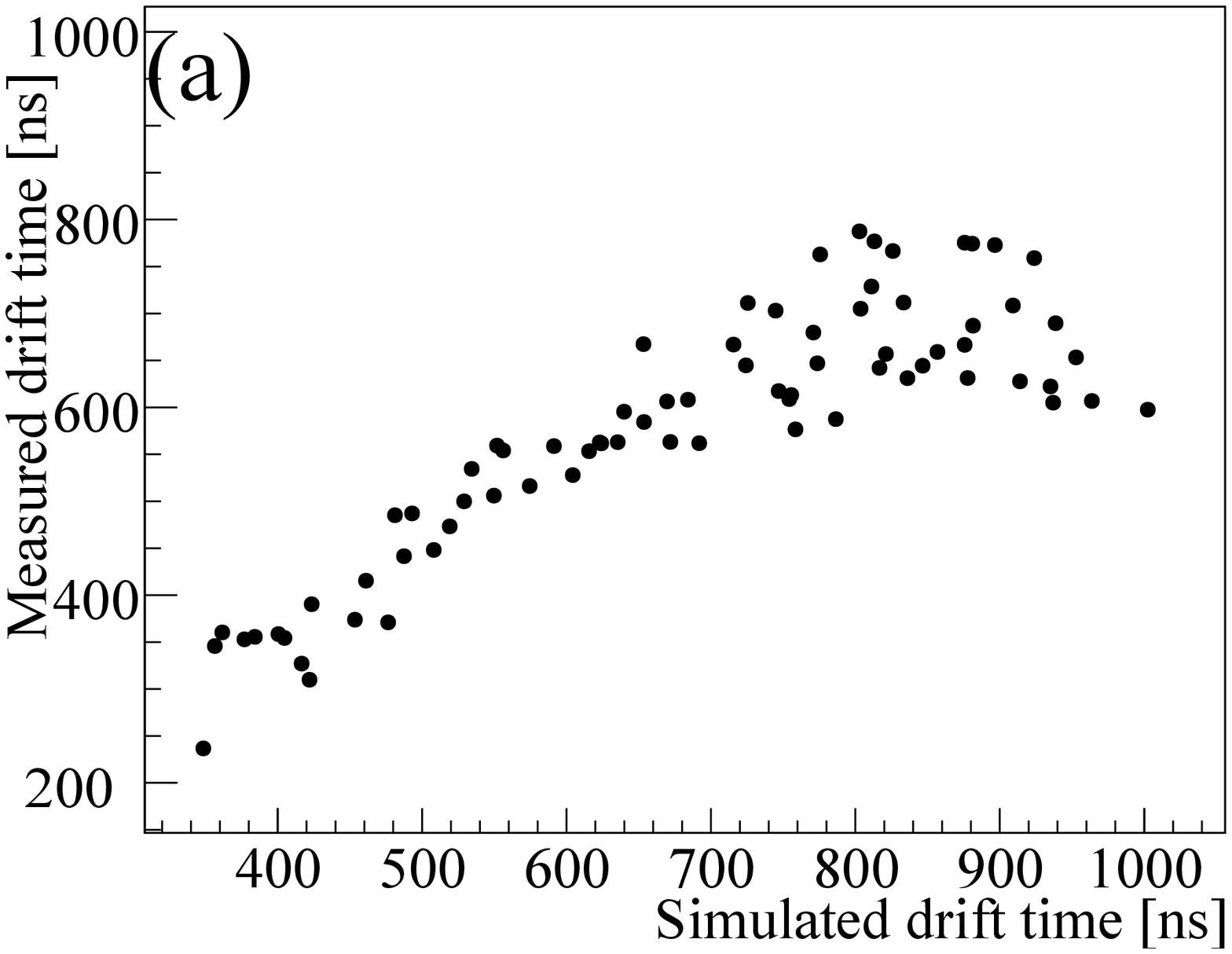}}
\subfigure{\includegraphics[width=0.475 \textwidth]{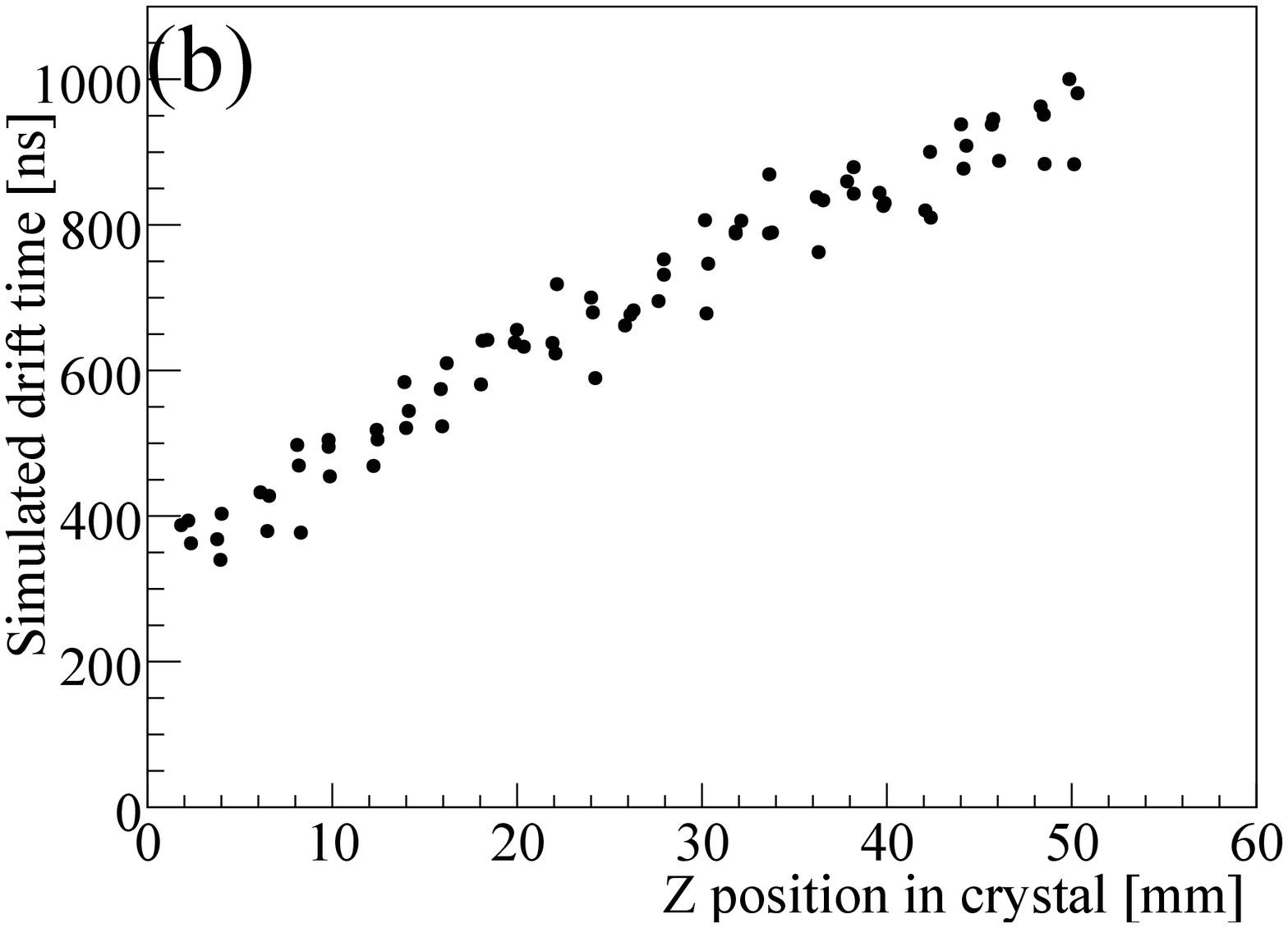}}
\subfigure{\includegraphics[width=0.45 \textwidth]{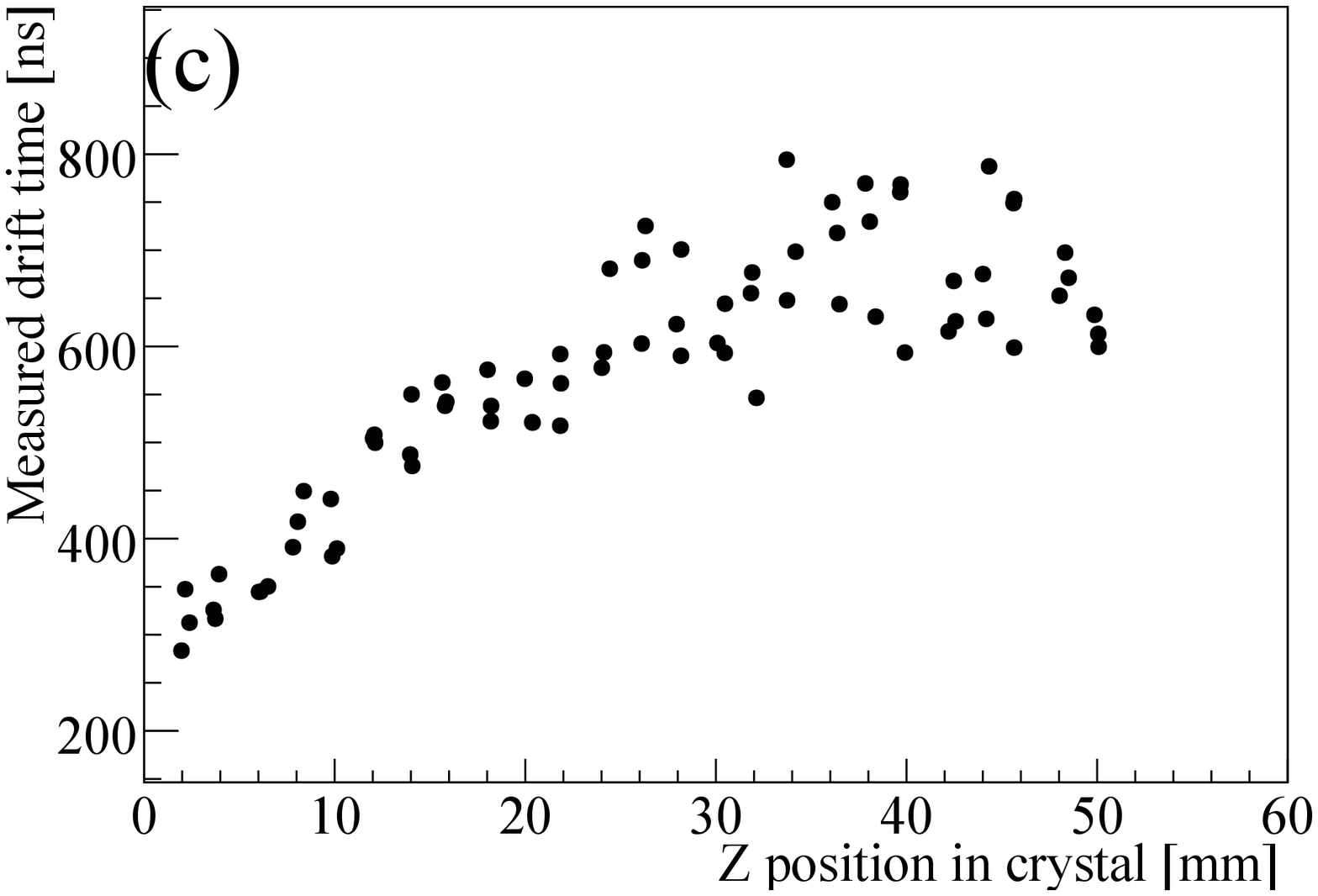}}
\caption{\label{fig:simpulse}Panel (a) shows the correlation between the measured drift time (y-axis) and actual drift time (x-axis) for simulated pulses with 1\% Gaussian noise (similar to the noise in 60\,keV events). Panel (b) shows how the actual drift time correlates with the Z-position of interactions in the crystal. Panel (c) shows how the measured drift time varies with position in the crystal. The measurement of drift time diverges from a linear relation at longer drift times; these events correspond to charge depositions far from the point contact and the full drift time become difficult to measure because of noise. About 35\% of events distributed uniformly in volume end up in this region.}
\end{figure}

Panel (a) of Figure \ref{fig:simpulse} shows the correlation between the drift time measured using the method described above and that obtained from the simulation for a set of simulated pulses with 1\% noise (relative to the amplitude of the pulses) distributed within the outer 5\,mm of the radius of the crystal. As the drift time becomes longer, the measured drift time diverges as the initial part of the pulse becomes difficult to distinguish from noise. Panel (b) shows that the simulated drift time is directly proportional to the z position of the charge depositions in the crystal, where $z=0$ corresponds to the point contact. Panel (c) shows how the measured drift time correlates with position in the crystal. The divergence starts for events occurring in the $\sim$35\% of the volume furthest from the point contact. For events in the remaining 65\% of the volume closest to the point contact, the measured drift time and that from the simulation are very well correlated. The data simulated with 0.1\% noise shows a smaller divergence since the noise has a smaller effect. Including pulses from the bulk of the crystal widens these distributions only slightly as the drift in the z direction contributes more to the total drift time than the drift in the radial direction.

\section{Data collection}
\label{sec:data}
\begin{figure}[!h]
   \centering
\includegraphics[width=0.7 \textwidth]{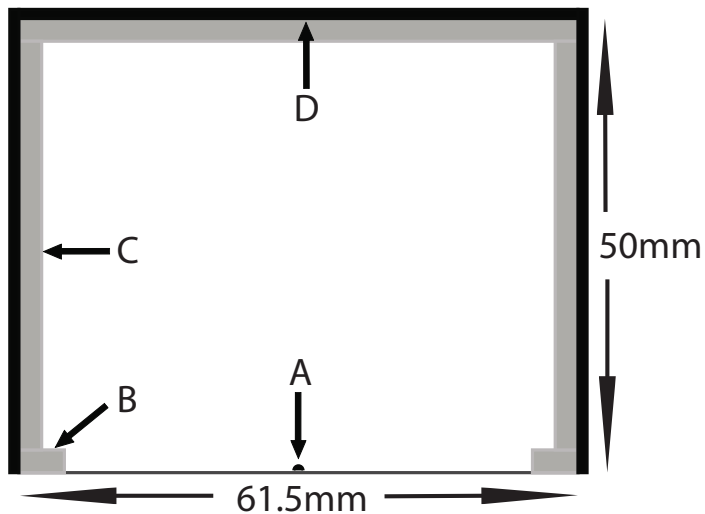}
\caption{\label{fig:PPCDetectorSide2} Side view schematic of PPC detector fabricated at LBNL, showing the point contact (A), the lithium-diffused n+ contact (B and C), and the vapor-deposited aluminum ohmic contact for the high voltage (D). The thickness of the lithium-diffused and aluminum layers are not to scale. The z axis referred to in the text points up with the origin at the point contact.}
\end{figure}

Data were collected with a custom PPC HPGe detector. The 61.5\,mm (dia.) by 50\,mm crystal was purchased from ORTEC \cite{OrtecWeb} and fabricated into a detector at Lawrence Berkeley National Laboratory (LBNL). The point contact is a hemispherical dimple with a diameter of 1\,mm $\pm$ 0.5\,mm. The crystal has a lithium-diffused n+ contact with aluminum metalization that covers the entire surface, except for the side with the point contact where the lithium-diffused region only wraps around for the outer 5\,mm and is not aluminized. A schematic is shown in Figure \ref{fig:PPCDetectorSide2}. The crystal was held in a prototype copper mount designed for the {\sc Majorana Demonstrator} experiment \cite{MJGeneral:2010i,Aalseth:2004yt,Elliott:2006i}. A custom, low-radioactivity electronic board with a FET that can be operated at cryogenic temperature was mounted close to the point contact, and the amplification loop was closed in a custom charge-sensitive preamplifier outside the cryostat. Processed pulses from the charge sensitive preamplifier were read out with a 16-bit 100\,MHz Struck SIS3302 (firmware version 1408) digitizing card. This detector and the associated electronics will be discussed in a later publication.

\section{Position dependence of drift time}

\begin{figure}[!h]
   \centering
\includegraphics[width=0.7 \textwidth]{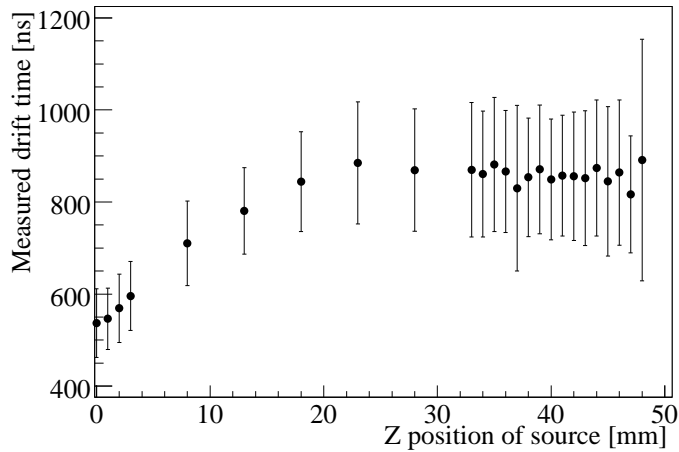}
\caption{\label{fig:scanYZdata_MeanDrift}Mean drift time of events as a function of z position of a collimated $^{241}$Am source. For each source position, the distribution of measured drift times of events between 58\,keV and 62\,keV was fit to a Gaussian. The mean and standard deviation from the fit were used to define the mean drift time and error as a function of z. The trend compares well to panel (c) of Figure \ref{fig:simpulse} which was done for the simulated pulses. }
\end{figure}

The drift time measurement was made using data from a collimated $^{241}$Am source directed at different axial positions (z axis) on the crystal. The $^{241}$Am source emits 59.5\,keV gamma rays that have a scattering length of order 1\,mm in germanium so that the position of the collimated source is well correlated with the position of the interactions in the crystal. Figure \ref{fig:scanYZdata_MeanDrift} shows the mean drift time of full energy events as a function of the source position. The drift time for each pulse was determined as $t_{90}-\tilde{t}_{0.1}$, using the algorithm from equation \ref{eqn:algo}, with a 500\,ns averaging time (50 samples for a 100\,MHz digitization). The mean drift time was obtained for each position of the source by fitting the distribution of drift time of many ($\sim 1000$) events between 58\,keV and 62\,keV using a Gaussian distribution. The error on the mean drift was taken as the standard deviation of the fitted Gaussian. The dependence of the mean drift time as a function of z is consistent with that observed in the simulated pulses. This indicates that, for the $\sim$ 30\,mm closest to the point contact, the measured drift time varies linearly with the axial position in the crystal. 

One can thus use the distribution of drift times from a detector to monitor the homogeneity of interactions in the crystal. This can be of use for low background experiments, where contaminated parts could be sources of background. An optimized arrangement of PPC crystals could, in principle, be used in an array to monitor the isotropy of external backgrounds. Additional sensing electrodes could be added to this type of detector to enhance the precision of the interaction sites in the crystal and lead to gamma ray tracking applications.

\section{Charge trapping model}

Following \cite{kephart_09} one can use the measured drift time of a cloud of charge carriers to examine the effects of their trapping by impurities in the crystal by implementing a simple model. The effective trapping length for this process can be modeled in the time domain as a ``charge-trapping time'', $\tau$, related to the charge carrier lifetime in the detector. As the drift time, $t_{drift}$, increases, the amount of charge in the cloud, $Q(t_{drift})$ decreases exponentially from the initial deposited charge, $Q_0$:

\begin{equation}
\label{eqn:drift}
Q(t_{drift}) = Q_0 e^{-\frac{t_{drift}}{\tau}}
\end{equation}
Since the energy deposited is proportional to the charge in the cloud, a loss of charge due to trapping results in an underestimate of the deposited energy if no correction is applied. Given the uncorrected measured energy, $E_{mes}$, and the charge trapping time, the corrected measure of the deposited energy, $E_{corr}$, can be recovered:

\begin{equation}
\label{eqn:energy}
E_{corr}(t_{drift}) = E_{mes}e^{\frac{t_{drift}}{\tau}}
\end{equation}
\section{Charge trapping characterization and energy resolution improvement}

In order to examine charge trapping effects, $^{60}$Co and $^{232}$Th source data were collected using the same detector setup as described above. Of particular interest is the double escape peak (DEP) at 1592\,keV from the $^{208}$Tl 2614\,keV line in the $^{232}$Th decay chain. Events in the DEP are intrinsically ``single-site events'' (SSE); that is, over 95\% of these events have only one energy deposition in the crystal and allow one to calibrate methods of rejecting multi-site events. Indeed, one would expect the charge trapping correction described above to work best for events having only 1 interaction in the crystal, as multi-site interactions would contain multiple components of drift time. Conversely, events from the single escape peak (SEP) at 2104\,keV are intrinsically multi-site.

 Multi-site events can effectively be rejected in PPC detectors using the ``A/E method'' \cite{Budjas:2009zu,Agostini:2010rd}. This method compares the maximum amplitude of the current pulse, $A$ (calculated using equation \ref{eqn:algo} with a shorter averaging time of 200\,ns), to the calibrated energy, $E$, of the event. In a multi-site event, the current pulse will have multiple peaks, whereas in a single-site interaction, the maximum amplitude of the current pulse will be proportional to the deposited energy. Charge trapping effects are examined with and without the A/E-based selection of single-site events. The crystal used in this experiment was shown to reject 93\% of multi-site events while preserving 90\% of single-site events using the A/E method. 
 
\begin{figure}[!h]
   \centering
\includegraphics[width=0.7 \textwidth]{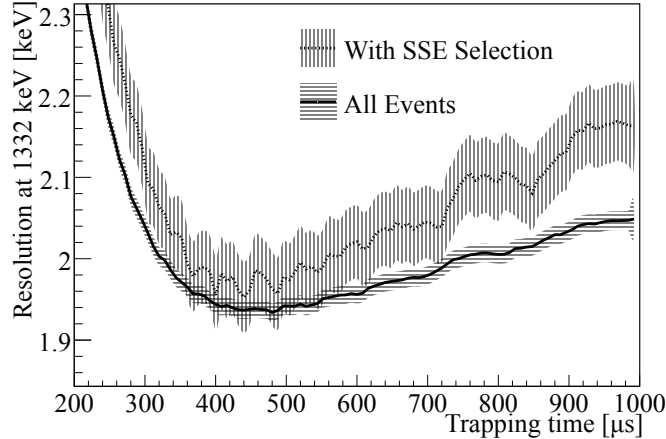}
\caption{\label{fig:tau_optimization_1173} Energy resolution of the $^{60}$Co 1332\,keV peak as a function of charge trapping time with SSE-selection applied (horizontal hash, light line) and for the raw spectrum (vertical hash, dark line). The region near the photopeak was fitted to a Gaussian with a linear background.  The standard deviation of the fit and its associated fit error were used to determine the FWHM and the resulting band. The error band for the data with SSE-selection applied is larger because of statistics (since most 1332\,keV events are multi-site and are thus removed). The presence of a global minimum (at $\tau \approx 450 \mu s$) is indicative of charge trapping in the crystal. As the charge trapping time increases, the resolution asymptotically tends to its uncorrected value.}
\end{figure}

Charge trapping effects in a detector can be characterized by applying equation \ref{eqn:energy} to the measured energy of events and determining the energy resolution as the charge trapping time, $\tau$, is varied. The drift time defined in section \ref{sec:math} is used to apply equation \ref{eqn:energy} for each event.  Figure \ref{fig:tau_optimization_1173} shows the full width at half maximum (FWHM) energy resolution of the 1332\,keV $^{60}$Co line as $\tau$ is varied, with and without the SSE-selection cut applied. Plotting these curves is a good tool for examining charge trapping effects in a crystal. If no charge trapping occurred, there would not be a global minimum since increasing the charge trapping time indefinitely is equivalent to applying no correction. The FWHM resolution with no correction is 2.30\,keV for the raw spectrum and 2.60\,keV for the SSE-selected spectrum; the asymptotic value is reached beyond the range plotted in the figure. Using these curves, the optimal charge trapping time is determined by fitting a parabola to the data in the region of the minimum. The uncertainty on $\tau$ is determined using the uncertainty on the energy resolution at the minimum of the fitted parabola; the range in $\tau$ where the parabola is within the uncertainty on the resolution is taken as the error on the determined value of $\tau$.

 It is interesting to note that, while the application of the single-site event selection cut degrades the energy resolution (from 2.30\,keV to 2.60\,keV, beyond the range shown in Figure \ref{fig:tau_optimization_1173} ), the charge trapping correction removes this effect. This can be explained by the fact that after the SSE events are selected, the sample of events contains a larger fraction of events with large drift times. The multi-site events with large measured drift times contain a significant contribution from the difference in the arrival times from multiple charge depositions rather than from actually long drift times. For this reason, a sample of events selected as single-site is more susceptible to charge trapping. This is an important effect to understand if one applies a single-site selection cut to remove backgrounds (such as in a neutrinoless double-beta decay experiment) and were to naively extrapolate the energy resolution that is measured using (predominantly) multiple-site events from calibration data.

\begin{figure}[!h]
   \centering
\includegraphics[width=0.7 \textwidth]{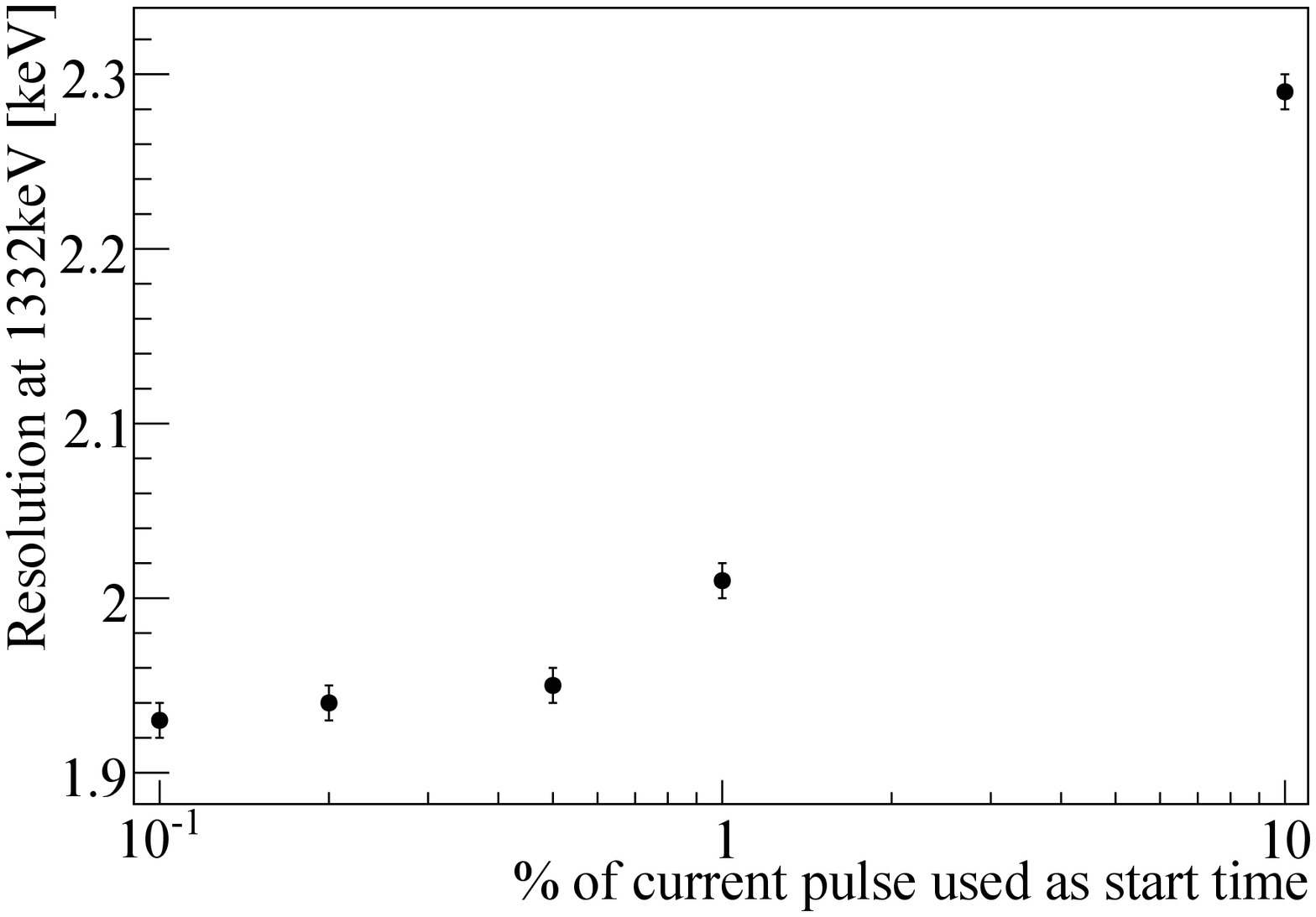}
\caption{\label{fig:Resolution1332VsTStartCurrentPercent} Energy resolution of the 1332\,keV peak as a function of the start time that was used to determined the beginning of the drift time, as measured in percent of the calculated average current waveform. The charge trapping correction is improved dramatically when using earlier start times.}
\end{figure}

 Figure \ref{fig:Resolution1332VsTStartCurrentPercent} shows the resolution of the 1332\,keV peak as different start times are used to define the drift time. There is a substantial improvement in energy resolution when using earlier start times to define the drift time. Incidentally, the start time defined using the 10\% point of the average current is similar to that obtained from using the 10\% rise of the charge pulse, as used in \cite{kephart_09}. This figure is indicative of the improvement that is obtained from using an optimized definition of drift time.

\begin{figure}[!ht]
   \centering
\subfigure{\includegraphics[width=0.45 \textwidth]{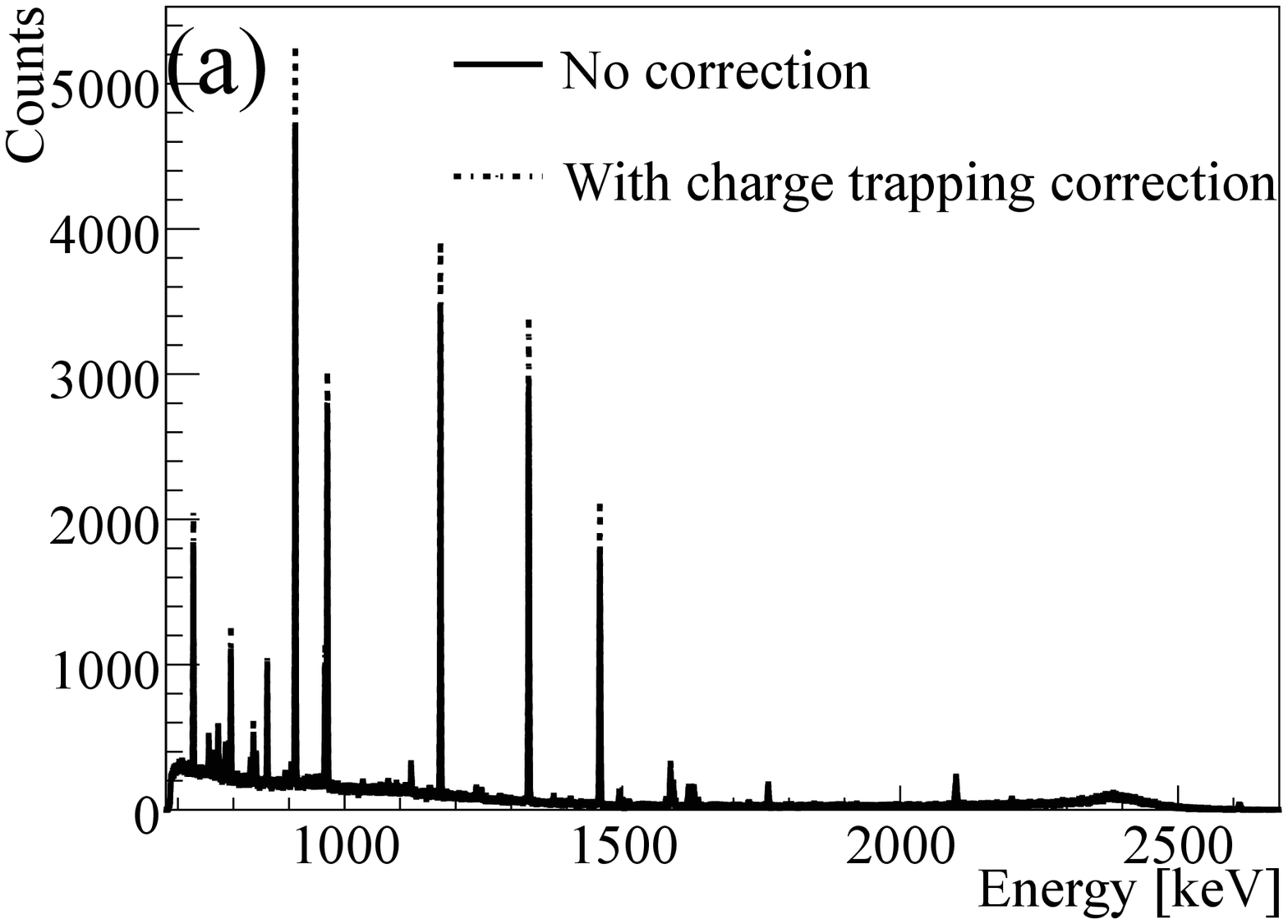}}
\subfigure{\includegraphics[width=0.48 \textwidth]{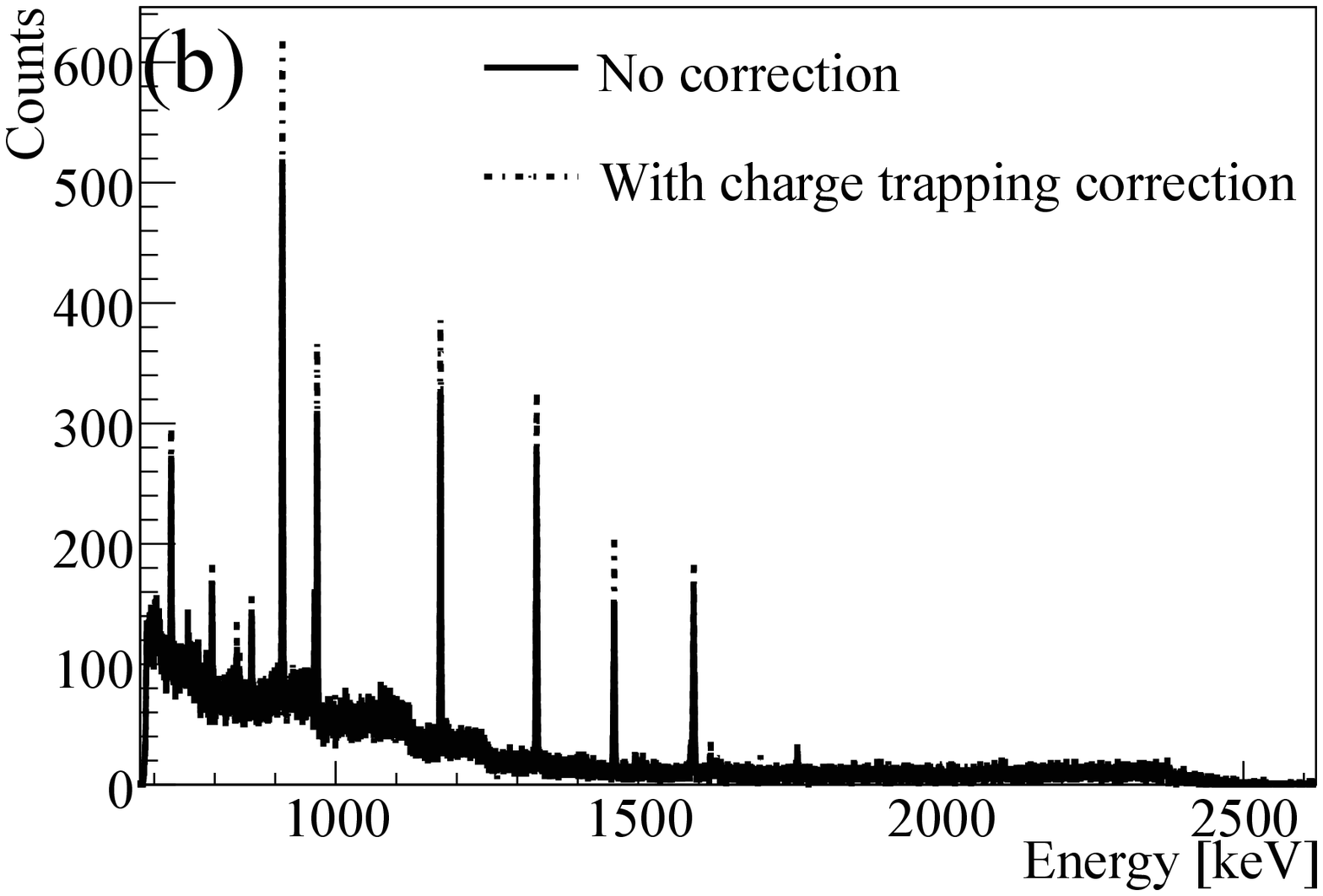}}
\subfigure{\includegraphics[width=0.45 \textwidth]{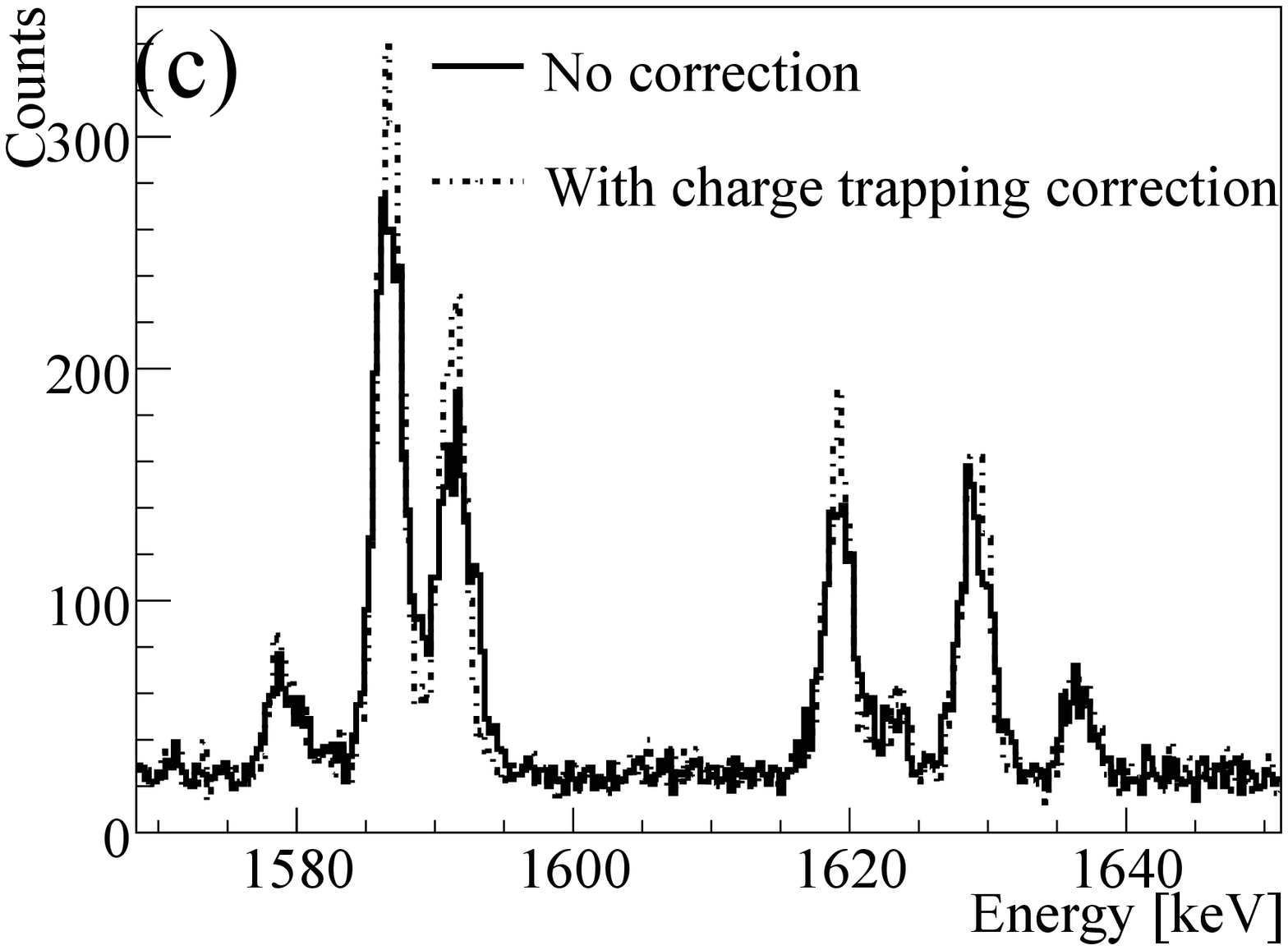}}
\subfigure{\includegraphics[width=0.48 \textwidth]{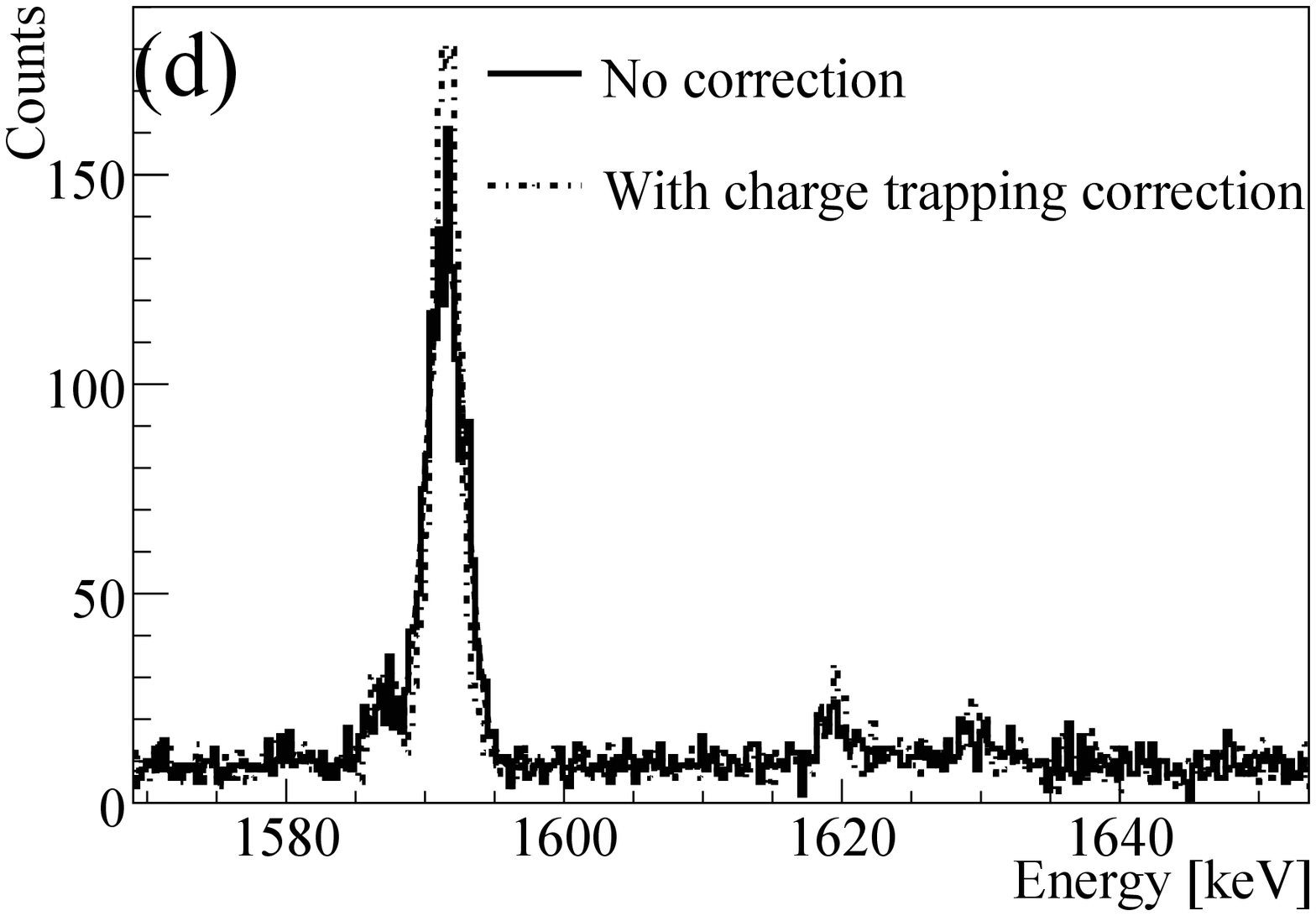}}

\caption{\label{fig:DEP_corrected} Energy spectrum of events before (solid) and after (dashed) a correction for charge trapping was applied. Panels (a) and (b) show a larger range of energies, whereas panels (c) and (d) highlight the region around the $^{208}$Tl DEP at 1592\,keV. Spectra are shown for all events (panels (a) and (c)) as well when the SSE-selection is applied (panels (b) and (d)). A change in energy scale from the charge trapping correction was accounted for in these figures.}
\end{figure}

\noindent Using peaks at 727\,keV, 911\,keV, 1173\,keV, 1332\,keV, 1592\,keV and 2104\,keV, the optimal charge trapping time and uncertainty were found for each energy (and tabulated in tables \ref{tab:ChargeTrapSummary} and \ref{tab:ChargeTrapSummarySSE}). An average charge trapping time was determined by weighing the value at each energy with its uncertainty and was then used to correct the entire energy spectrum. The average charge trapping time for this crystal was determined to be $\approx$ 460$\mu$s. Figure \ref{fig:DEP_corrected} shows the energy spectrum before (solid) and after (dashed) the charge trapping correction was applied, for events with and without the SSE-selections. Panels (a) and (b) show the energy spectrum over a large range, whereas panels (c) and (d) show the spectrum in the region around the $^{208}$Tl DEP, highlighting the effect of the SSE-selection. In particular, one notes the substantial reduction in the $^{228}$Ac peak at 1588\,keV. The charge trapping correction introduces a change in energy scale that was accounted for in these figures; it was verified that this change in energy scale was linear (see equation \ref{eqn:energy}).

 The improvement in energy resolution from the charge trapping correction is summarized in Tables \ref{tab:ChargeTrapSummary} and \ref{tab:ChargeTrapSummarySSE} for all events and for those selected as single-site, respectively. The results presented here show that the charge trapping correction originally suggested by \cite{kephart_09} greatly benefits from an optimal determination of the drift time. For all energies, the charge trapping correction removes the bias in energy resolution introduced from applying the SSE-selection; the corrected energy resolution is now consistent between all events and those selected as single-site. The data presented here show that charge trapping corrections can be important for $^{76}$Ge neutrinoless double-beta decay experiments that strive to have good energy resolution while implementing cuts to remove multiple-site interactions.

\begin{table*}[!ht]
   \centering
   \footnotesize
 	\begin{tabular}{lllll|ll}
      \textbf{Energy} & \textbf{Isotope} & \textbf{Nominal} & \textbf{Best}  & \textbf{Optimal} & \textbf{FWHM at}& \textbf{$\Delta \%$}\\
        &  &  \textbf{FWHM} & \textbf{FWHM}& \textbf{$\tau$} &  \textbf{$\tau$=451\,$\mu$s} & \\            
      \hline
      727\,keV  &$^{228}$Ac & 1.56(2)\,keV  & 1.39(2)\,keV & 451(104)\,$\mu$s & 1.41(2)\,keV & 10\%\\
      911\,keV  &$^{228}$Ac & 1.76(1)\,keV  & 1.53(1)\,keV & 471(108)\,$\mu$s & 1.56(1)\,keV & 11\%\\
      1173\,keV  &$^{60}$Co & 2.13(1)\,keV  & 1.81(1)\,keV & 454(88)\,$\mu$s & 1.84(1)\,keV & 14\%\\     
      1332\,keV  &$^{60}$Co & 2.30(1)\,keV  & 1.93(1)\,keV & 470(84)\,$\mu$s & 1.96(1)\,keV & 14\%\\     
      1592\,keV  &$^{208}$Tl DEP & 3.30(12)\,keV  & 2.13(1)\,keV & 437(52)\,$\mu$s & 2.21(7)\,keV & 33\%\\   
      2104\,keV  &$^{208}$Tl SEP & 3.96(11)\,keV  & 3.51(9)\,keV & 508(392)\,$\mu$s& 3.55(9)\,keV & 11\%\\             
       \end{tabular}
\caption{ \label{tab:ChargeTrapSummary}Summary of charge trapping correction performance for selected lines from Figure \ref{fig:DEP_corrected}a (no SSE-selection applied). The table shows the nominal FWHM energy resolution (with no correction), the best resolution found by varying $\tau$, the value of $\tau$ that gives the best resolution, the resolution at the average value of $\tau$ and the relative improvement in energy resolution at the average value of $\tau$ for various lines in the energy spectrum for $^{60}$Co  and isotopes in the $^{232}$Th decay chain. Uncertainties are shown in parentheses.}
\end{table*}

\begin{table*}[!ht]
   \centering
    \footnotesize
 	\begin{tabular}{lllll|ll} 
      \textbf{Energy} & \textbf{Isotope} & \textbf{Nominal} & \textbf{Best}  & \textbf{Optimal} & \textbf{FWHM at}& \textbf{$\Delta \%$}\\
        &  &  \textbf{FWHM} & \textbf{FWHM}& \textbf{$\tau$} &  \textbf{$\tau$=464\,$\mu$s} & \\
      \hline 
      727\,keV  &$^{228}$Ac & 1.69(9)\,keV  & 1.44(7)\,keV & 515(164)\,$\mu$s & 1.46(7)\,keV & 14\%\\
      911\,keV  &$^{228}$Ac & 1.92(4)\,keV  & 1.58(3)\,keV & 519(216)\,$\mu$s & 1.65(3)\,keV & 14\%\\
      1173\,keV  &$^{60}$Co & 2.24(5)\,keV  & 1.77(4)\,keV & 489(128)\,$\mu$s & 1.82(4)\,keV & 19\%\\     
      1332\,keV  &$^{60}$Co & 2.59(6)\,keV  & 1.95(4)\,keV & 452(100)\,$\mu$s & 2.01(5)\,keV & 23\%\\     
      1592\,keV  &$^{208}$Tl DEP & 3.26(10)\,keV  & 2.19(4)\,keV & 432(96)\,$\mu$s & 2.28(6)\,keV & 30\%\\               
      \end{tabular}
\caption{ \label{tab:ChargeTrapSummarySSE}Summary of charge trapping correction performance for selected lines from Figure \ref{fig:DEP_corrected}b (events selected as single-site). The table shows the nominal FWHM energy resolution (with no correction), the best resolution found by varying $\tau$, the value of $\tau$ that gives the best resolution, the resolution at the average value of $\tau$ and the relative improvement in energy resolution at the average value of $\tau$ for various lines in the energy spectrum for $^{60}$Co  and isotopes in the $^{232}$Th decay chain. Uncertainties are shown in parentheses. The $^{208}$Tl single escape peak is not included since the SSE cut removes essentially all of those multi-site events.}
\end{table*}

\section{Conclusion}
With PPC detectors gaining popularity as the detector of choice in various types of low background experiments using germanium, understanding their properties has become an area of increased interest. This work presented a method to estimate the drift time of charge carriers in PPC detectors. Given the drift time, one has some sensitivity in determining the position of an energy deposition, which can be of use for monitoring the isotropy of events in a low background detector. It has also been shown, following previous work by \cite{kephart_09}, that one can use the drift time to look for evidence of charge trapping in PPC detectors. A correction to improve the energy resolution of PPC detectors was found to benefit greatly from using an optimized measure of drift time. It was noted that selecting single-site events using the A/E method can result in an apparent increase in charge trapping effects and a biased estimate of the energy resolution. Since the energy resolution has a direct impact on the sensitivity of neutrinoless double-beta decay experimental searches, these types of experiments must correctly account for charge trapping effects when determining their energy resolution if they use an A/E-type selection to reduce backgrounds.

\section{Acknowledgments}
The authors would like to thank the {\sc Majorana} Collaboration for valuable input on this work, as well as Justin Tang and Barbara Wang for their help with the initial charge trapping studies. This work was supported by the U.S. Department of Energy under Contract No. DE-AC02-05CH11231.

\bibliography{MJReferences}
\bibliographystyle{model1-num-names}

\bigskip{\small \smallskip\noindent Updated \today.}
\end{document}